\documentclass[preprint,12pt,authoryear]{elsarticle}

\usepackage{amssymb}
\usepackage{amsmath}
\usepackage{lipsum}
\usepackage{url}
\usepackage[normalem]{ulem}
\usepackage{subcaption} 

\begin{document}
\begin{frontmatter}

\title{Studying Ionospheric Phase Structure Functions Using Wide-Band uGMRT (Band-4) Interferometric Data}

\author[]{Dipanjan Banerjee$^{1}$ \corref{cor1}}
\ead{dipanjanbanerjee20@gmail.com}
\address{Department of Physics, Banwarilal Bhalotia College, Asansol, WB, Pin: 713303, India}

\author[]{Abhik Ghosh$^{2}$ \corref{cor1}}
\ead{abhik.physicist@gmail.com}
\address{Department of Physics, Banwarilal Bhalotia College, Asansol, WB, Pin: 713303, India}

\author[]{Sushanta K. Mondal$^{3}$}
\ead{skm.phy@skbu.ac.in}
\address{Department of Physics, Sidho-Kanho-Birsha University, Ranchi Road, Purulia 723104, India}

\author[]{Parimal Ghosh$^{4}$}
\ead{parimal34@gmail.com}
\address{Department of Physics, Banwarilal Bhalotia College, Asansol, WB, Pin: 713303, India}

\cortext[cor1]{These authors contributed equally to this work.}

\begin{abstract}
Interferometric observations of the low-frequency radio sky ($<$1~GHz) are largely limited by systematic effects introduced by the ionosphere. Here, we analyse a ten-hour nighttime uGMRT Band-4 observation of 3C48 to characterise ionospheric phase fluctuations across baselines up to 25~km. We compute spatial phase structure functions across three sub-bands (575--725~MHz), revealing power-law behaviour consistent with turbulence and a diffractive scale $r_\mathrm{diff} \approx 6.7 - 8.3$~km useful for assessing calibration requirements. The turbulence exhibits anisotropy with smallest scales perpendicular to Earth's magnetic field - consistent with wave-like structures such as MSTIDs rather than field-aligned irregularities. These findings from a single case study demonstrate uGMRT's sensitivity for ionospheric characterisation at low-latitudes ($\sim19^\circ$N) and inform direction-dependent calibration strategies for similar conditions.
\end{abstract}


\end{frontmatter}

\section{Introduction}
\label{introduction}
At low radio frequencies, the ionosphere is the main contributor to phase disruptions, affecting signals from astronomical sources \citep{Taylor99}. A monochromatic interferometer detects the phase shift caused by differences in signal delay as the wavefront passes through the ionosphere at two separate locations. These phase shifts are due to variations in the mean refractive index along the signal paths reaching each antenna.

As long as propagation remains in the refractive regime, the interferometer's phase differences can be interpreted as differences in excess path length the additional distance light appears to travel due to the ionosphere between the two ray paths. This assumes that signals from the source travel nearly parallel through the ionosphere to each antenna. 
In the diffractive regime, however, wavefronts lose their parallel alignment, and intensity fluctuations can arise not only between antennas but even across a single antenna’s surface. In such cases, calibration methods based on refractive assumptions break down \citep{Vedantham15, Gasperin19}.

The excess path length results from the deviation of the ionospheric refractive index from unity, integrated along the geometric signal path. This effect depends on the column density of free electrons, known as the Total Electron Content (TEC), along the line of sight.

A useful statistical tool for characterizing ionospheric phase irregularities is the \textit{spatial phase structure function}, which describes how phase fluctuations vary with antenna separation across an array. It quantifies the average squared difference in interferometric phase between pairs of antennas as a function of their relative spacing. This function captures how phase variability changes with scale and provides insight into the spatial behaviour of ionospheric disturbances.

In turbulent media like the ionosphere, the structure function often follows a \textit{power-law relationship} with antenna separation. The exponent in this power law, typically denoted as $\beta$, reveals the nature of the underlying turbulence. A value of $\beta = 5/3$ is expected for fully developed isotropic turbulence, following Kolmogorov theory. However, deviations from this value can indicate the presence of other effects, such as organized wave structures or anisotropic conditions in the ionosphere \citep{Spencer55, Mevius16, Jordan17, Loi15}.

Although ionosondes and GPS-based methods provide valuable measurements of large-scale ionospheric properties \citep{Arora16, Gasperin18, Fallows20}, radio interferometers offer unprecedented resolution for probing ionospheric structures on finer spatial and temporal scales. Arrays such as Low-Frequency Array (LOFAR), Murchison Widefield Array (MWA), Very Large Array (VLA), and more recently, Square Kilometre Array (SKA) pathfinders, have been used to examine ionospheric behaviour with high precision. These studies have revealed rich ionospheric dynamics, including anisotropy aligned with the Earth's magnetic field, spectral signatures of Travelling Ionospheric Disturbances (TIDs), and turbulent cascades across a wide range of spatial scales \citep{Rufenach1972-bk,Vanvelthoven90,Helmboldt12,Helmboldt12b,Loi15}.

The upgraded Giant Metrewave Radio Telescope (uGMRT), situated near Pune in India is a powerful but still underused tool for studying the ionosphere \citep{Swarup1991, Gupta17}. It has thirty large 45-meter dishes and can observe a broad spectrum of radio frequencies (from 120 to 1460 MHz) using five different frequency bands. Thanks to its improved receivers and advanced processing systems, the uGMRT can measure ionospheric phase changes with high sensitivity and detail across both time and frequency. Its location in a low-latitude region is also important, as it allows scientists to study ionospheric conditions that are different from those found near the poles or the equator \citep{Mangla22, Mangla23}.

In this study, we take a close look at the ionospheric phase structure function derived from a single ten-hour Band-4 observation conducted with the upgraded Giant Metrewave Radio Telescope (uGMRT; \citep{Gupta17}). The observation targets one of the brightest flux calibrators in the sky, 3C48. 3C48 is a compact, distant quasar (flux $\sim$ 28 Jy near Band-4 central frequency of 650 MHz) often used as a flux and bandpass calibrator in radio astronomy \citep{Scaife12, Perley13}. It is compact and very bright, with a stable signal across a broad range of radio frequencies, which makes it especially useful for studying ionospheric effects.

The paper is structured as follows: Section~\ref{sec:observation} outlines the 
observational data and phase processing. Section~\ref{sec:strf} presents 
ionospheric spatial variability and the two-dimensional structure function. 
Here we show 1D and 2D results from the first sub-band (575--600~MHz), 
comparing baseline orientations to the local projected magnetic field to 
investigate anisotropy in phase fluctuations \citep{Loi15, Mevius16}. 
Section~\ref{sec:discussion} provides discussion and conclusions. 
Similar results from the other two sub-bands are shown in Appendix~A.

\section{Observation and Processing}
\label{sec:observation}
The uGMRT is a sensitive radio interferometer optimized for low-frequency observations. It comprises 30 parabolic antennas, each with a diameter of 45 meters, distributed over a 25 km region. Fourteen antennas are located within a central $1.4 \times 1.4$ km$^2$ square, and the remaining 16 are spread along three 14 km-long arms forming a Y-shaped layout. This Y-shaped configuration enables high-resolution imaging and multi-directional sampling of ionospheric irregularities. Located between the magnetic equator and the northern crest of the equatorial ionization anomaly (EIA), the GMRT is ideally positioned for ionospheric studies \citep{Mangla22}.

We conducted 10 hours of nighttime observations of the bright quasar 3C48 using the uGMRT at Band-4 (project code: $47\_003$ \footnote{\url{https://naps.ncra.tifr.res.in/goa/data/search}}) in November 2024. This observation occurred during quiet geomagnetic conditions  ($K_{\rm p}$ index < 3) and high solar activity was reported\footnote{\url{https://www.noaa.gov/}} (F10.7 $\sim$ 200 sfu). The data span a 200 MHz bandwidth (550–750 MHz) with a central frequency of 650 MHz, using 8192 channels at 24.41 kHz resolution and an integration time of 2.68 seconds. For ease of computation and to avoid frequency ranges significantly affected by radio frequency interference (RFI), we selected a subset of three relatively clean sub-bands: 575–600 MHz, 600–625 MHz, and 700–725 MHz. We found overall good agreement among the analysed bands. More details about the observations are summarized in Table~\ref{obs_sum}.

RFIs was mitigated using a two-step process. Initial strong RFI was manually flagged, followed by automated flagging with the \texttt{TFCROP} algorithm from the CASA package \citep{CASA22}. This algorithm identifies statistical outliers in time-frequency space using thresholds of 4$\sigma$ (time) and 3$\sigma$ (frequency), with the \texttt{EXTEND} mode enabled to flag data if more than 50\% of bins are contaminated. Approximately 30\% of the data for both RR and LL polarizations was flagged, with additional iterative flagging applied based on calibration solutions.

\begin{table}[h]
    \centering
    \caption{{\bf Observation summary}}
    \label{obs_sum}
    \begin{tabular}{l c}
        \hline
        Observation ID  & 47\_003 \\
        Observation Date & 23 November 2024 \\
        Observation Window (IST) & 5:21 PM to 3:27 AM \\
         Processed Frequency Bands & 575–600, 600–625, and 700–725 MHz \\
        $K_{\rm p}$ index & $<$ 3 \\
        F10.7 & $\sim$ 200 sfu \\
        Channel Width & 24.41 kHz \\
        Integration Time & 2.68 seconds \\
        Measured Correlations & RR and LL \\
        Functional Antennas & 30 \\
        Target Source & 3C48 \\
        Right Ascension (J2000) & 01$^h$37$^m$41.301596$^s$ \\
        Declination (J2000) & +33$^\circ$09$'$35.25459$''$ \\
        Total Exposure Time & 10 h \\
        \hline
    \end{tabular}
\end{table}

\subsection{Calibration and Phase Processing}
After flagging, standard calibration is performed using conventional CASA tasks with 3C48 as the flux, bandpass, and phase calibrator. Flux density was set using the \citet{Scaife12} model via the CASA task \texttt{SETJY}. Calibration began with delay corrections using \texttt{GAINCAL} (\texttt{gaintype=K}) relative to a reference antenna (C06), followed by bandpass calibration using \texttt{BANDPASS} with a solution interval of 10 seconds and 8 channels ($\sim$194 kHz). Although both amplitude and phase are calibrated, we focus here on the phase solutions for ionospheric studies. A total of $\sim 32\%$ of the data was flagged post-calibration and RFI mitigation.

Phase differences between antennas were modelled as contributions from ionospheric, instrumental, source structure, and phase ambiguity effects \citep{Helmboldt12, Mangla22}. Source-related phase contributions were removed using model visibilities, while $2\pi$ ambiguities were addressed through phase unwrapping. Outliers and phase jumps were detected and flagged using a Local Outlier Factor (LOF) algorithm combined with a cosine filter \citep{Helmboldt12, Banerjee25}. Flagged time steps with abrupt phase changes (exceeding $5\sigma$) were interpolated to maintain temporal continuity. Further details regarding the phase correction process can be found in \citealt{Banerjee25}.

Short-timescale ($<$ 1~hr) ionospheric fluctuations were isolated by subtracting a 1-hour boxcar-smoothed instrumental phase trend from the unwrapped phase time series. Although absolute instrumental phases could not be fully separated, this approach enabled analysis of short-timescale TEC variations ($<$ 1 hour). Residual phase variations post-continuum subtraction were typically within $\pm0.5$ radians for central square antennas and up to $\pm5.0$ radians for arm antennas \citep{Banerjee25}. The observation on 23 November 2024 occurred during relatively quiet geomagnetic conditions ($K_{\rm p}$ < 3), while the solar radio flux was elevated (F10.7 $\sim$ 200 sfu). During the observation we detected a period of rapid phase and amplitude variations between approximately 19:30 and 21:45 IST, which degraded data quality. The interval from 20:15--20:55 IST was severely affected and was therefore flagged and excluded from the analysis. The exact cause of this instability is uncertain and could be related to RFI contamination or instrumental effects \citep{Banerjee25}.

\section{Results: Ionospheric \textbf{Phase Structure Function}}
\label{sec:strf}

The phase structure function is a key tool in characterizing ionospheric turbulence. It encapsulates the scale-dependent nature of phase fluctuations due to turbulent plasma density variations. For Kolmogorov-type turbulence, both the phase and refractive index structure functions exhibit power-law behaviour, and are tightly linked to their respective power spectra. These formulations are essential in modelling ionospheric effects on radio wave propagation, particularly for applications in radio astronomy, satellite communication, and GNSS positioning \citep{Van2009PhDT}.

The spatial characteristics of a turbulent medium like the ionosphere are usually described using its power spectrum, or alternatively its phase correlation function, which corresponds to the Fourier transform of the power spectrum. However, in practice, it is often more practical to describe how the ionospheric phase changes across space using the phase structure function \citep{Mevius16}, 

\begin{equation}\label{eq:phasestructure}
\Xi(r) = \left\langle \left( \phi(r') - \phi(r'+r) \right)^2 \right\rangle,
\end{equation}

where \( \phi(r) \) represents the ionospheric phase at position \( r \), and the angle brackets denote an ensemble average. This function measures the expected squared difference in phase between two antenna locations separated by a distance \( r \).

In the case of Kolmogorov turbulence, which describes a wide range of turbulent media including the ionosphere, the phase structure function follows a power-law behaviour within the inertial range:

\begin{equation}\label{eq:1dKolmogorov}
\Xi(r) = \left( \frac{r}{r_{\rm diff}} \right)^\beta,
\end{equation}

where \( r_{\rm diff} \) denotes the \emph{diffractive scale}, the spatial separation at which the phase variance reaches \(1~\mathrm{rad}^2\), and \( \beta = 5/3 \) corresponds to pure Kolmogorov turbulence \citep{Narayan92}. This scale encapsulates the strength of the turbulence: smaller values of \( r_{\rm diff} \) indicate stronger phase fluctuations over smaller spatial separations.
\begin{figure}[ht]
    \centering    
    \includegraphics[width=1\linewidth]{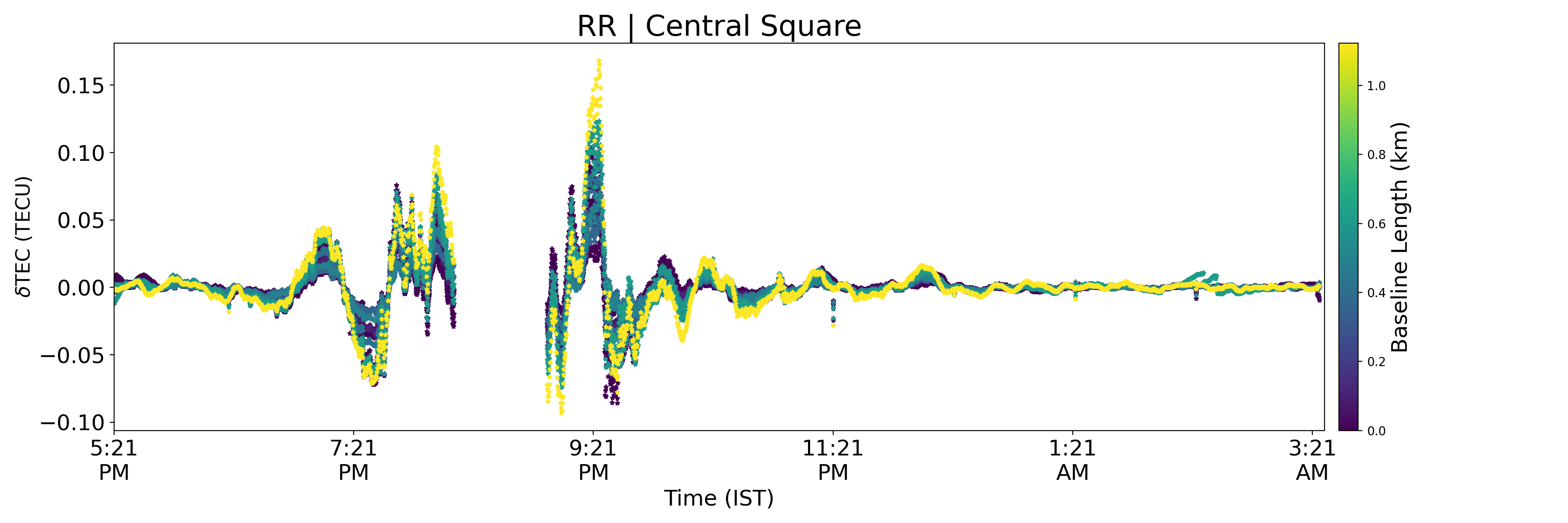}
    \includegraphics[width=1\linewidth]{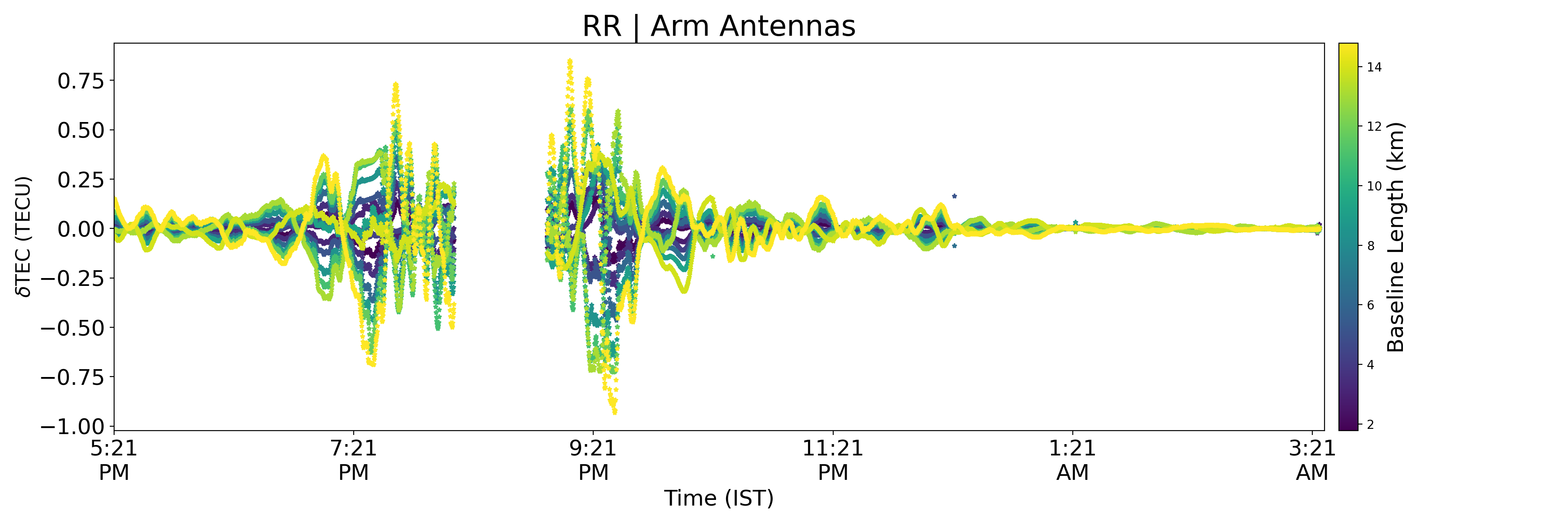}    
    \caption{Differential TEC ($\delta {\rm TEC}$) for the RR polarization of uGMRT baselines relative to the reference antenna `C06' as a function of time. The colour bar indicates baseline length. (Top row) Baselines within the central square. (Bottom row) Extended antenna arm baselines.}
    \label{fig:dtec_base}
\end{figure}

\begin{figure}
    \centering
    \includegraphics[width=1\linewidth]{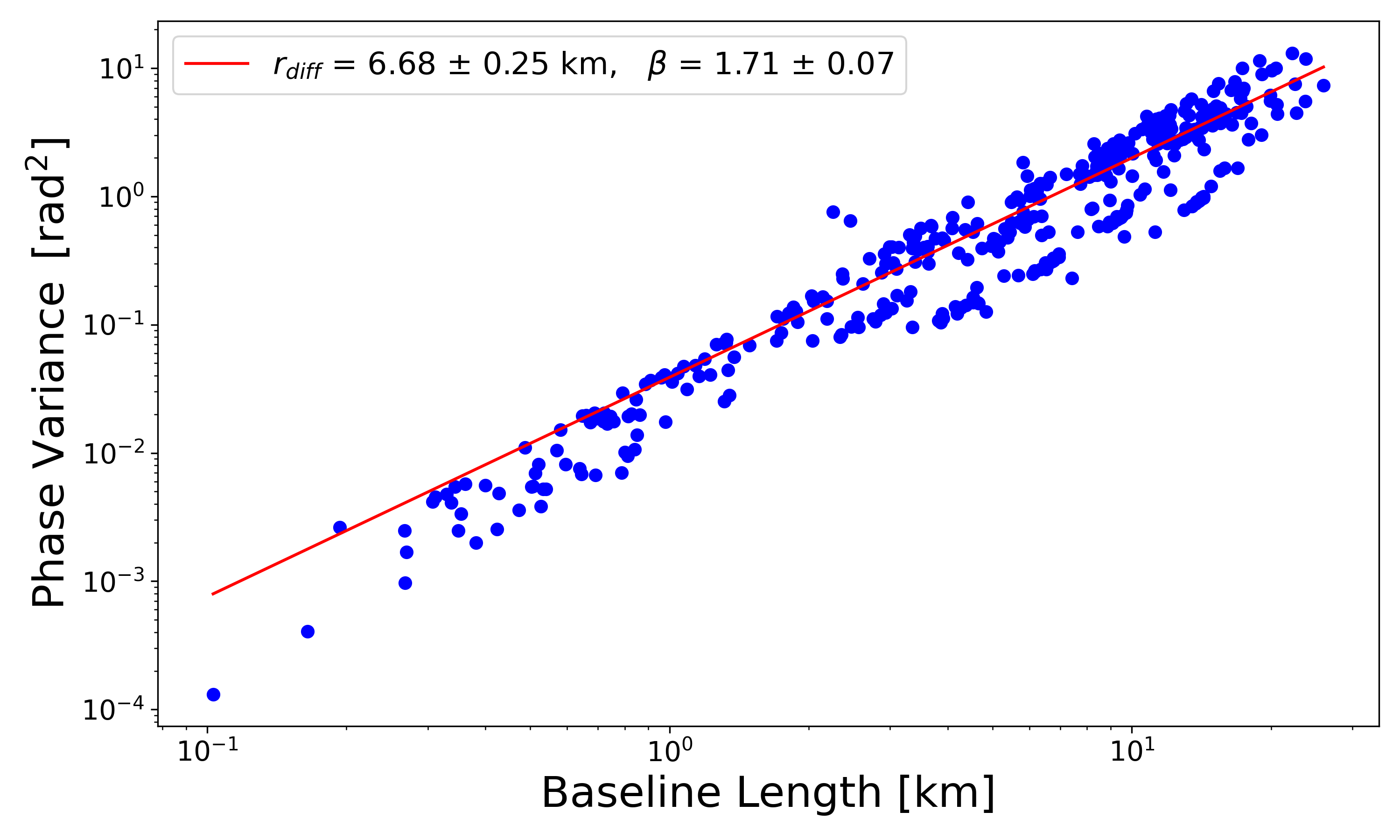}
    \caption{This figure shows the phase structure function at 600 MHz. The blue points represent the measured phase variance as a function of baseline length, while the red solid line shows the fitted one-dimensional power-law model.}
    \label{fig:strf1d}
\end{figure}

Differential phases from Section~\ref{sec:observation} yield slant-path $\delta$TEC, which is subsequently converted to vertical TEC (vTEC) with a thin-layer approximation at peak height of $\sim 300$ km using IRI model\footnote{\url{https://www.ionolab.org/iriplasonline/}} \citep{Helmboldt12,Banerjee25}. 
The vTEC values (Figure~\ref{fig:dtec_base}) are then converted back to phase delays using the representative sub-band frequency (e.g., at 600 MHz in Figure~\ref{fig:strf1d}) through Equation \eqref{eq:phasedelay} for phase structure function analysis.

\begin{equation}
\Delta \phi_{\text{ion}} \approx 8.45 \left( \frac{1\,\text{GHz}}{\nu} \right) \left( \frac{\delta \text{TEC}}{1\,\text{TECU}} \right) \quad \text{radians}
\label{eq:phasedelay}
\end{equation}

The extended 10-hour observation, combined with ionospheric motion relative to the array, provides sufficient independent samples to invoke the ergodic theorem for statistical averaging. Figure \ref{fig:strf1d} also shows an example of the spatial un-binned phase structure function for a typical Band-4 uGMRT night-time observation.

The structure function displays a clear power-law trend across a broad range of baseline lengths, from roughly 100 meters up to 25 kilometres. There is no sign of it flattening out at the longest baselines (around 20 km), which would typically suggest the outer scale of turbulence - where the structure function is anticipated to stop increasing and become constant. To determine the outer scale of turbulence more definitively, we would need to include much longer  baselines \citep{Mevius16}, which we currently don't have access to. There is ongoing effort to extend uGMRT baselines upto 100 km \citep{Patra13},  this can enhance the angular resolution of the uGMRT by a factor of 5 and also help us to determine the outer scale of turbulence. At short baselines ($<100$~m), the structure function can flatten. This effect, observed in LOFAR's 29-night 3C196 study \citep{Mevius16}, arises from a noise 
floor. We observe no flattening for baselines $>100$~m.

Figure~\ref{fig:strf1d} shows the measured phase structure function along with the best-fit 1D model. The results follow a power-law with an index of \( \beta = 1.71 \pm 0.07 \), which is slightly steeper than the expected Kolmogorov value of $1.66$. This suggests that there may be additional structures in the ionosphere affecting the measurements \citep{Vanvelthoven90, Loi15}. We estimated the diffractive scale at 600 MHz is \( r_{\rm diff} \sim 6.68 \)~km, which is the distance where the phase variance reaches 1~rad\(^2\). 

A small diffractive scale means that the phase of the signal changes a lot across the field of view. When the diffractive scale is small ($<$ 5 km), the signal also stays coherent for a shorter time, and this depends on the length of the baseline \citep{Vedantham15}. To correct for these fast-changing phase variations, calibration must be done in many directions and at short time intervals \citep{Gasperin19}. However, the number of directions and how often we can update the calibration depends on how bright the sources are and how much independent data is available \citep{Intema09, Bregman12}. Overall, \( r_{\rm diff} \) is a useful indicator of how stable the ionosphere is for radio imaging.

\begin{figure}[ht]
    \centering
    \includegraphics[width=1\linewidth]{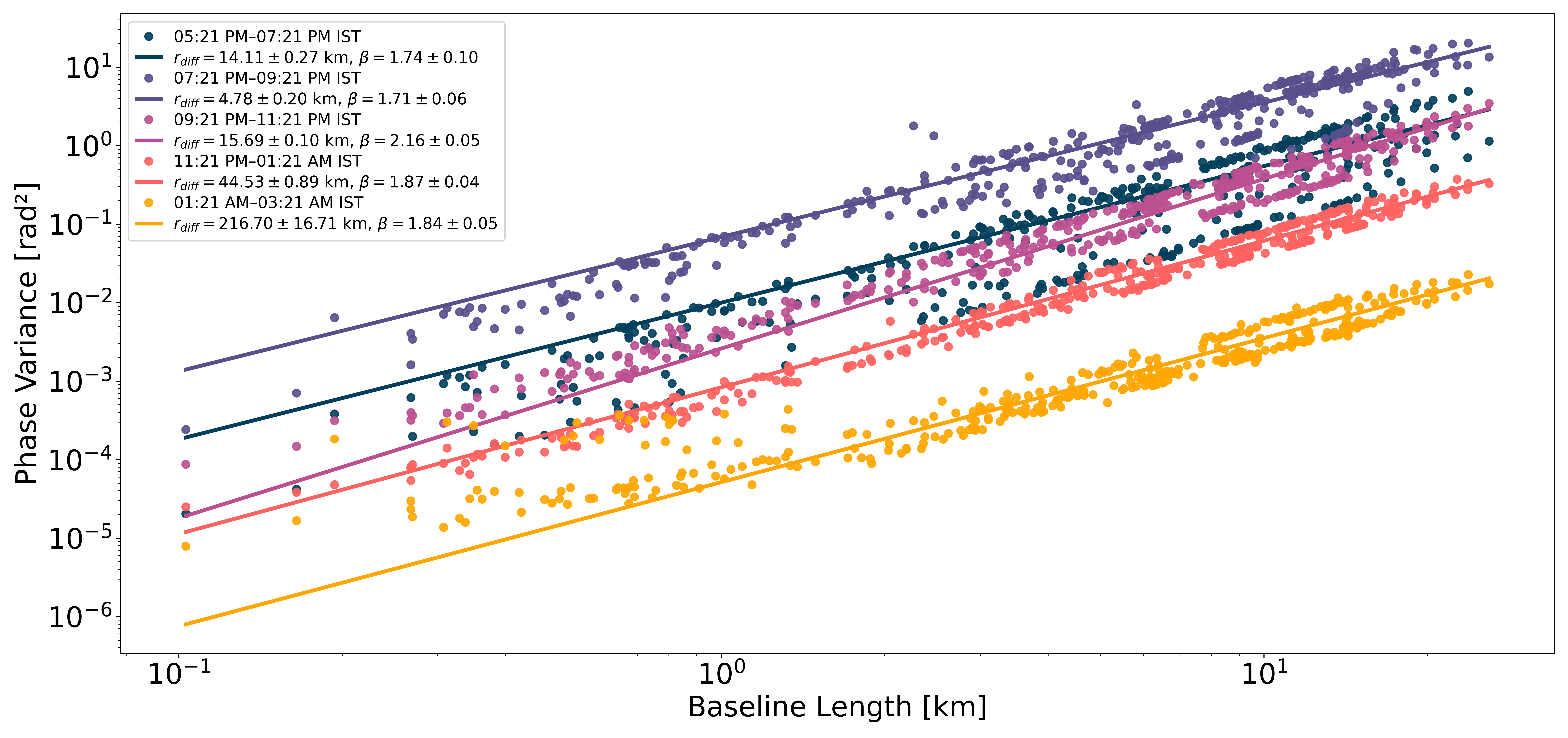}
    \caption{The figure shows the phase structure function divided into five time chunks of 2 hour each. The initial part of the observation is visibly more affected by ionospheric disturbances, as reflected by the lower ${\rm r_{diff}}$ values. This trend is also observed in differential TEC values in Figure~\ref{fig:dtec_base}.}
    \label{fig:strfchuncked}
\end{figure}

Earth rotation causes the projected baseline length and orientation of a fixed antenna pair to evolve during the scan on 3C48. To capture the ionospheric variation within shorter time scales we also repeated our analysis within $\sim$ 2 hour time window (\citet{Gasperin18}, Figure~\ref{fig:strfchuncked}). We find the power law slope is mostly similar, even though the diffractive length scale changes mostly from low to higher values, which suggest a quieter ionosphere as our observation progresses. A similar trend also can be seen in Figure~\ref{fig:dtec_base}, where the variation in  $\delta {\rm TEC}$ values are more rapid towards the start of the observation. Together, these plots provide insights into how the ionospheric irregularities evolve across Band-4 frequencies.

\subsection{2D Structure Function}
In Figure \ref{fig:strf1d} we notice a band-like pattern in the phase structure function. We find that the \textit{scale of the structure function} (i.e., the $r_{\text{diff}}$ value) varies with direction, even though the \textit{slope} remains roughly the same. This indicates that the \textit{ionospheric irregularities are anisotropic}, meaning their properties differ depending on direction.

The observed pattern may also be attributed to the presence of \textit{large, wave-like structures} within the ionosphere. These waves appear to propagate in the direction where the \textit{smallest diffractive scales} are found. The anisotropy of the ionosphere has been reported in multiple earlier studies \citep{Wheelon01, Spencer55, Singleton70, Mevius16}.

\begin{figure}[ht]
    \centering
    \includegraphics[width=1\linewidth]{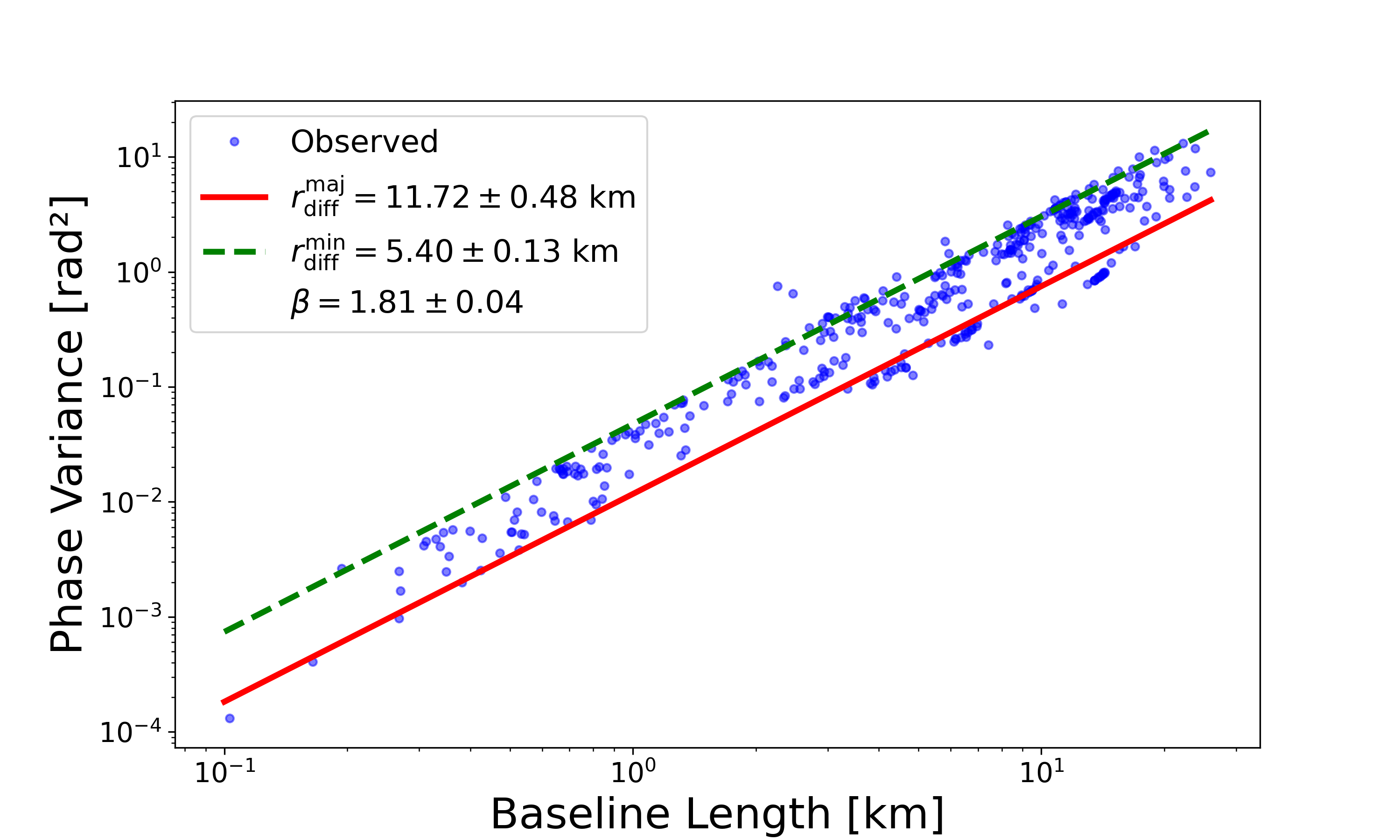}
    \caption{Two-dimensional phase structure function at 600 MHz. The green dotted and red solid lines represent the major and minor axis projections, respectively, based on the 2D structure function model described in Equation~\ref{eq:2DKolmogorov}.}
    \label{fig:strf2d}
\end{figure}

To account for this anisotropy, the structure function can be generalized to a \textit{two-dimensional} form that allows for direction-dependent variations (Figure \ref{fig:strf2d}). The anisotropic structure function is defined as,

\begin{equation}
    \Xi(\mathbf{r}) = \left( \mathbf{r}^\top R^\top C R \mathbf{r} \right)^{\beta/2},
\label{eq:2DKolmogorov}
\end{equation}

where $R$ denotes the $2 \times 2$ rotation matrix that aligns the coordinate system with the principal axes of anisotropy:
\[
R = 
\begin{bmatrix}
    \cos\alpha & -\sin\alpha \\
    \sin\alpha & \cos\alpha
\end{bmatrix},
\quad
C = \mathrm{diag}\left(\frac{1}{r_{\mathrm{diff,maj}}^2}, \frac{1}{r_{\mathrm{diff,min}}^2}\right).
\]
Here, $\mathbf{r} = (r_x, r_y)$ be the baseline vector, where $r_x$ and $r_y$ are its components along the east-west (EW) and north-south (NS) directions, respectively. The diffractive scales $r_{\mathrm{diff,maj}}$ and $r_{\mathrm{diff,min}}$ are defined with respect to an orthogonal coordinate system aligned with these directions. $\alpha$ is the angle of the major axis measured from geographic east, and $\beta$ is the power-law exponent as in the isotropic case.

In Figure~\ref{fig:strf2d}, we present the 2D structure function, where the lines indicate the projections along the major and minor axes of the fitted model described in Equation~\ref{eq:2DKolmogorov}. The best-fit parameters are $r_{\mathrm{diff,maj}} = 11.7$~km and 
$r_{\mathrm{diff,min}} = 5.4$~km, yielding a moderate anisotropy ratio of $2.17$. 
The major axis orientation is $\alpha = 135^\circ$, indicating that the 
irregularities are elongated along the SE--NW direction. This orientation is 
not aligned with the projected geomagnetic field direction, suggesting that 
the observed structures are consistent with wave-like disturbances such as MSTIDs  \citep{Vanvelthoven90}.

\subsection{Anisotropy in the ionospheric structures}
Previous studies \citep{Spencer55,Singleton70,Wheelon01, Loi15, Mevius16} found that irregularities in the ionosphere tend to stretch along Earth’s magnetic field lines. This alignment is particularly expected in the case of field-aligned irregularities or elongated plasma structures, which can be guided by the geomagnetic field. However, such alignment is not universal and may be disrupted by large-scale wave-like disturbances such as TIDs \citep{Mevius16}.

To investigate this uneven anisotropy, we computed the Earth's magnetic field vector at the uGMRT site for the date of our observation (23rd November, 2024) using the World Magnetic Model \citep{Chulliat2020wmm}. The magnetic field was evaluated at an altitude of approximately 300 km, which is a typical average height for peak ionospheric electron density according to the IRI model. Since the angle between the baselines and the magnetic field lines changes over time, we can't just calculate one value of variation for each baseline. Instead, we grouped the data by both angle and length. 

\begin{figure}[ht]
    \centering
    \includegraphics[width=1\linewidth]{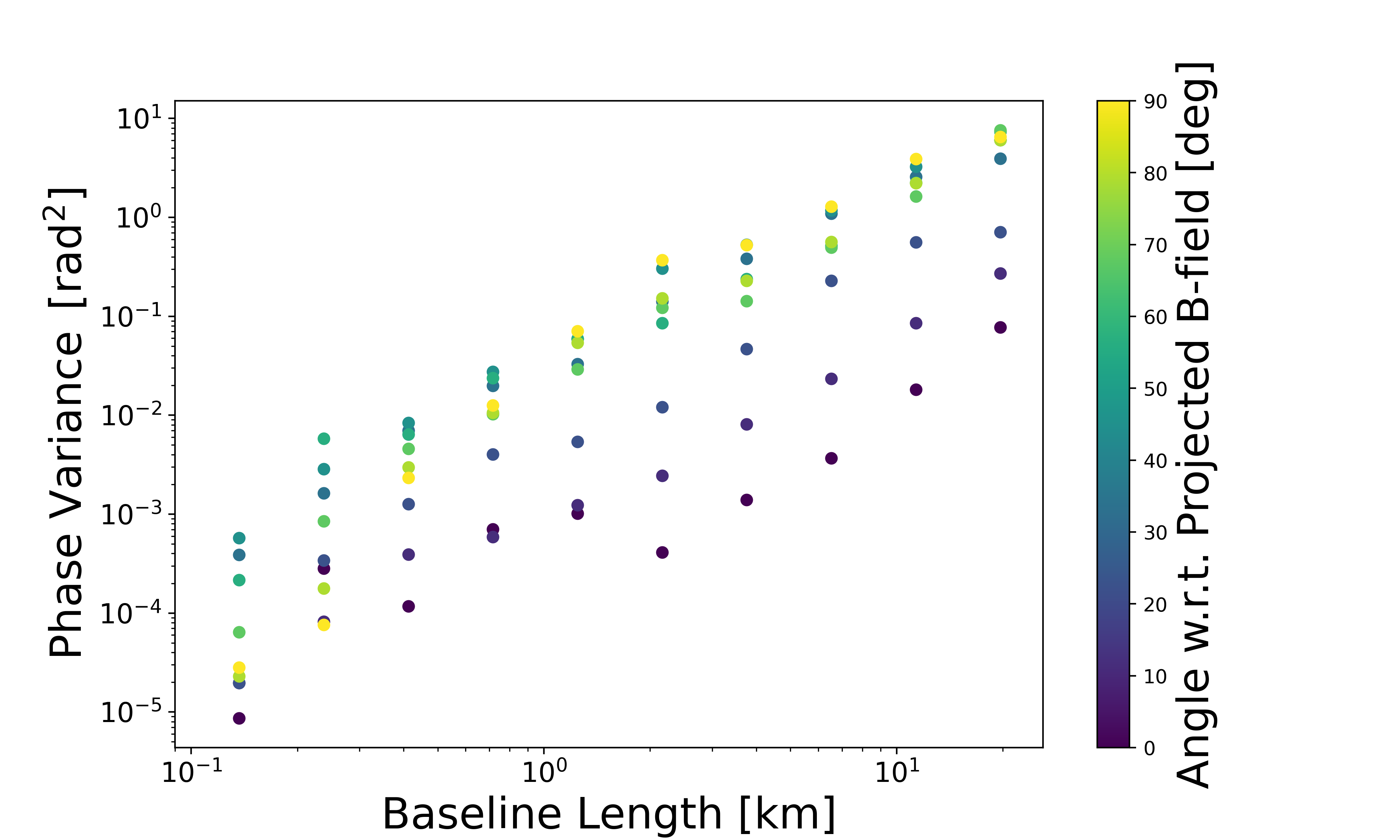}
    \caption{The figure shows the phase structure function where the data is binned by the angle relative to the projected Earth's magnetic field. The colour bar indicates the angle in degrees.}
    \label{fig:all_angles}
\end{figure}

Figure~\ref{fig:all_angles} displays the phase structure function, with the phase variance data grouped based on the angle relative to the projected Earth's magnetic field. The angle between each baseline vector and the local magnetic field was determined using $\cos(\theta) = \mathbf{b} \cdot \mathbf{B} / |\mathbf{b}||\mathbf{B}|$,
where $\mathbf{b}$ represent the baseline vector and $\mathbf{B}$ denotes the magnetic field vector.

We notice the phase variance varies for each fixed baseline length when the angle between the baselines and projected magnetic field changes. 
For the lowest and highest angle bins this variation is close to two orders of magnitude in the phase variance (Figure~\ref{fig:all_angles}). 

\begin{figure}
    \centering
    \includegraphics[width=0.49\linewidth]{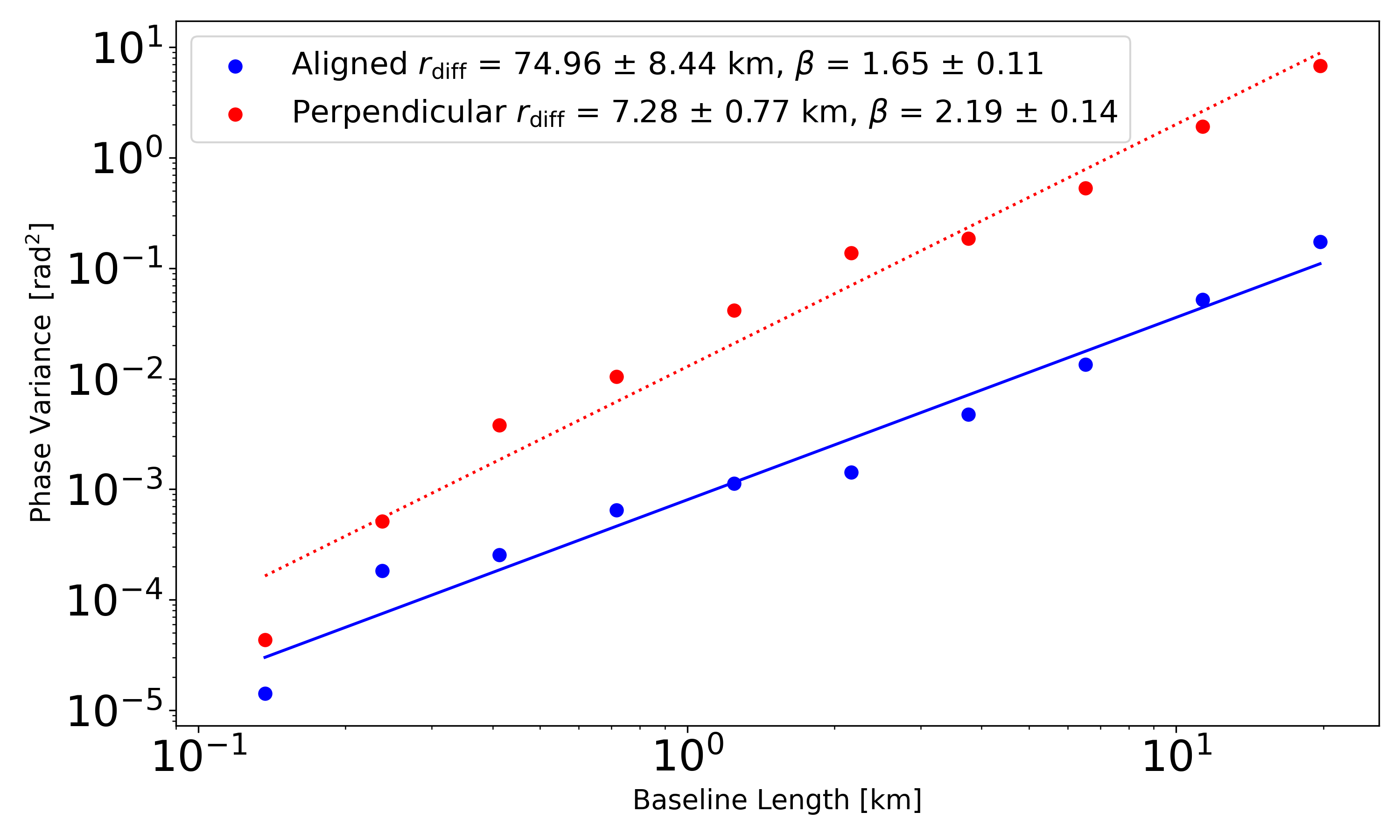}
    \includegraphics[width=0.49\linewidth]{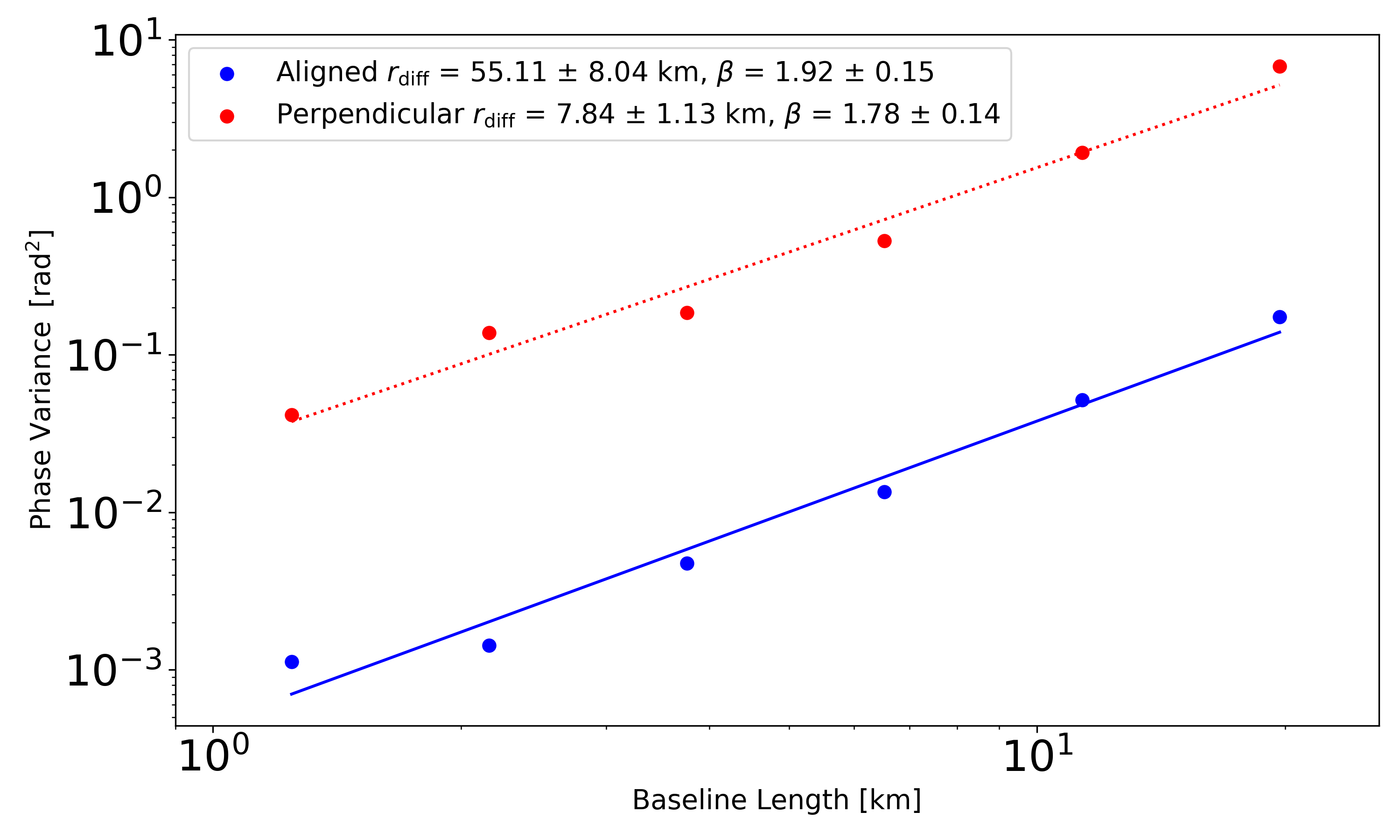}
    \caption{Left: Phase structure function comparing two groups of baselines based on their angle to the Earth's magnetic field: those nearly aligned (between $0^\circ$ and $20^\circ$, shown with blue solid lines) and those nearly perpendicular (between $70^\circ$ and $90^\circ$, shown with red dotted lines). Each line shows a 1D power-law fit of phase variance with baseline length; Right: Using only baselines longer than 1~km. The phase variance shows reduced scatter, resulting in a more reliable power-law fit.}
    \label{fig:ap}
\end{figure}

To explore this pattern more closely in our observation, we focused on two specific angle ranges: $0^\circ$--$20^\circ$, which is approximately aligned with the magnetic field, and $70^\circ$--$90^\circ$, which is nearly perpendicular to it. 
The 1D structure function model fits (Eqn. \ref{eq:1dKolmogorov}) show the diffractive scale along the field lines is much greater than the perpendicular direction (Figure \ref{fig:ap}, left). This suggest the ionospheric structures in our observation is not filed aligned. If we choose a subset of baselines above 1 km (Figure \ref{fig:ap}, right) the model fit improves (Residual Standard Error (RSE) is reduced by $\sim 70\%$ and $\sim 43\%$ for the aligned and perpendicular fits, respectively) but our conclusion remain unchanged. The steep power-law slope ($\beta \sim 2.0$) and B-field misalignment are characteristic of MSTID wave-like structures \citep{Vanvelthoven90}. Note, here we used 1D projections (instead of the full 2D model) as they provide stable fits for binned angular subsets ($0$--$20^{\circ}$ aligned and $70$--$90^{\circ}$ perpendicular), focusing on differences in slope and $r_{\mathrm{diff}}$. This choice enhances interpretability for anisotropy trends and aligns with prior work \citep{Mevius16}. The full 2D model (Equation~\ref{eq:2DKolmogorov}) was applied to the unbinned data (Figure~\ref{fig:strf2d}).  We also tested 2D fits for the angular subsets ($0^\circ$--$20^\circ$, and $70^\circ$--$90^\circ$). Although the fits are noisier due to scattered phase variance (particularly for field-aligned baselines), the results are consistent with the 1D analysis.

\section{Summary and Conclusions}
\label{sec:discussion}
The analysis presented here uses uGMRT Band 4 data, closely following the phase structure function methodology of \citet{Mevius16}, but applied to a distinct frequency regime and instrument. This work extends phase structure function studies to a higher frequency range compared to LOFAR's 150 MHz observations. 

The differing ranges of observing frequencies have important consequences. Ionospheric phase fluctuations scale inversely with frequency, resulting in smaller phase RMS and often smaller outer scales at higher frequencies for identical ionospheric conditions. Consequently, phase stability tends to improve swiftly as frequency increases, affecting the required calibration intervals and the characteristics of direction-dependent errors.

In this study, we analysed nighttime ionospheric phase fluctuations using a 10-hour observation of the bright flux calibrator source 3C48 with the uGMRT. Specifically, we studied the ionospheric phase structure function at around 575-725 MHz while tracking the bright radio source 3C48. This function helps us understand how the phase of a radio signal changes with distance, due to turbulence in the ionosphere. 

We estimated differential TEC between antenna pairs from unwrapped interferometric phase differences after removing the slow continuum variation with a 1-hour boxcar filter. The corrected phase differences were converted to TEC differences and corresponding phase delays. Using these values, we calculated the phase structure function as a function of baseline length, which follows a power-law trend. The fitted slope is slightly steeper than the Kolmogorov expectation, indicating possible additional small-scale ionospheric structures or more complex turbulence.

Further, we extended our analysis of the ionospheric phase structure function to two dimensions, allowing us to investigate directional effects. While the slope of the structure function (\( \beta \)) remains approximately constant across different directions, the diffractive scale \( r_{\text{diff}} \) shows clear directional dependence. This indicates that the ionospheric turbulence is \textit{anisotropic}.

To model this anisotropy, we fitted the structure function using a 2D power-law model. 2D analysis yields a moderate anisotropy ratio of $2.17$. The major axis orientation is $\alpha = 135^\circ$, indicating that the irregularities are elongated along the SE--NW direction. This orientation differs from the projected geomagnetic field direction and is consistent with wave-like ionospheric disturbances such as MSTIDs \citep{Vanvelthoven90}. Additionally, we compared 1D structure functions for baselines nearly aligned (0--20$^\circ$) vs perpendicular (70--90$^\circ$) to Earth's magnetic field, finding $r_\mathrm{diff}$ larger along the field than across it (Fig.~\ref{fig:ap}). This indicates stronger phase fluctuations perpendicular to field lines, with more stable fits and slopes $\sim2$ (steeper than Kolmogorov 5/3) for baselines $>$ 1~km (Figure~\ref{fig:ap}). For radio interferometric calibration, the difference in signal behaviour along and across ionospheric structures needs to be considered when calculating noise levels. Additionally, ionospheric models used for calibration such as SPAM \citep{Intema09}  could be improved by incorporating the fact that these structures tend to follow the magnetic field and are tilted relative to the Earth's surface \citep{Mevius16}.

This case study provides empirical phase structure function measurements 
at intermediate frequencies (575--725~MHz), complementing lower-frequency 
LOFAR results and offering context for mid-frequency SKA operations. 
We note this analysis uses only nighttime observations from 23 November 
2024 (winter season), so results do not account for diurnal or seasonal 
variations. These findings demonstrate uGMRT's potential for ionospheric 
studies from its low-latitude location near the geomagnetic equator 
\citep{OPIO20151640}.

\section*{Acknowledgements}
We thank the staff of the GMRT for their support in making these observations possible. The GMRT is operated by the National Centre for Radio Astrophysics (NCRA) of the Tata Institute of Fundamental Research (TIFR). The authors utilized ChatGPT for AI-assisted copy editing and improving the manuscript's language. AG would like to thank IUCAA, Pune, for providing support through the associateship programme, including access to the computational facility at IUCAA.

All authors acknowledge the financial support received through the SERB-SURE grant (SUR/2022/000595) from the Science and Engineering Research Board, a statutory body under the Department of Science and Technology (DST), Government of India. We would like to sincerely thank the reviewers for their thoughtful and constructive feedback on our manuscript.

\appendix

\section{Sub-Band Variations in Ionospheric Phase Structure Function (Band-4)}
\label{append1}

\begin{figure}[ht]
    \centering
    \includegraphics[width=0.49\linewidth]{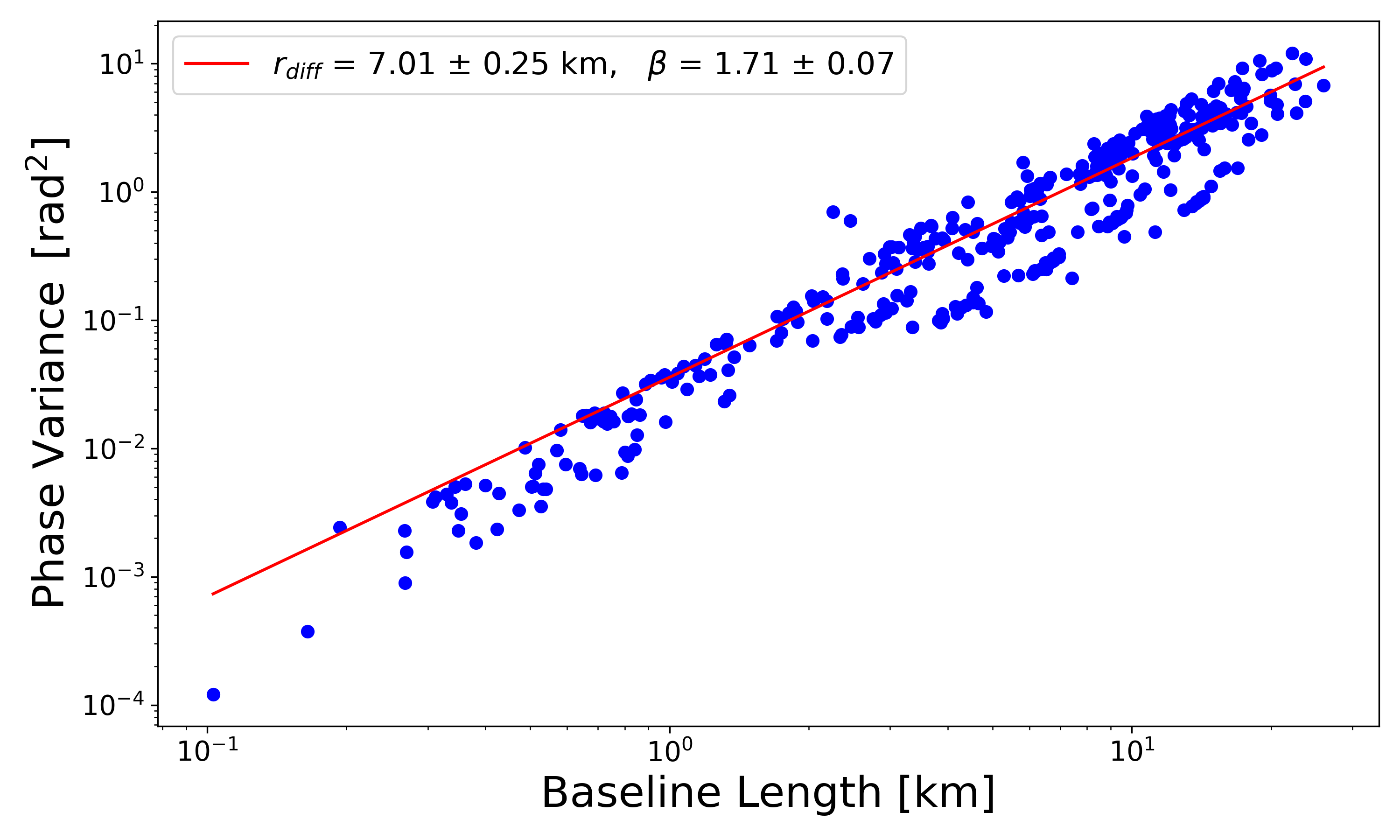}
    \hfill
    \includegraphics[width=0.49\linewidth]{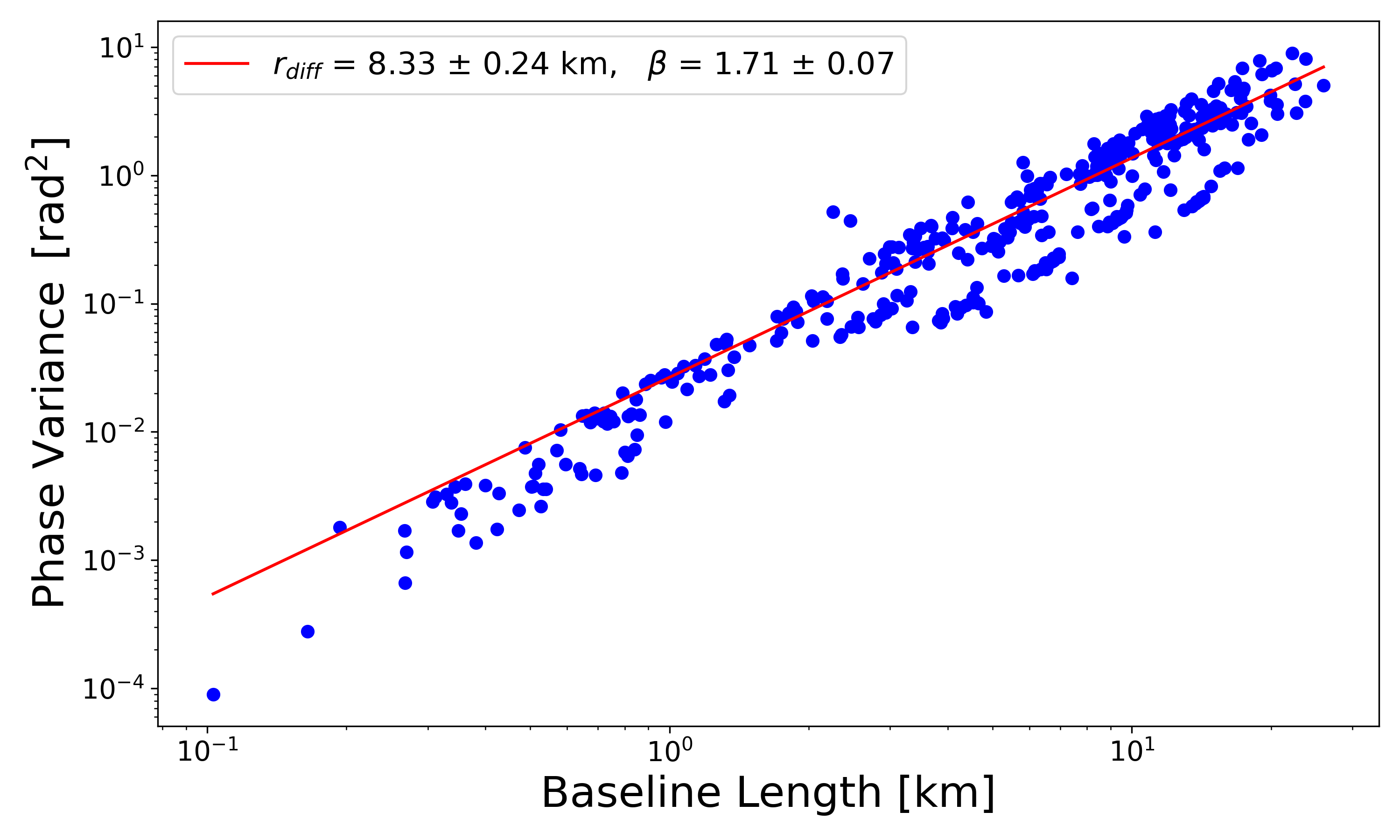}
    \caption{Phase structure function as a function of baseline length. Left: At 625 MHz. Right: At 725 MHz. Blue points show the observed phase variance, and the red lines indicate the fitted power-law model.}
    \label{fig:structure_func_comp}
\end{figure}

\begin{figure}[ht]
    \centering
    \includegraphics[width=0.49\linewidth]{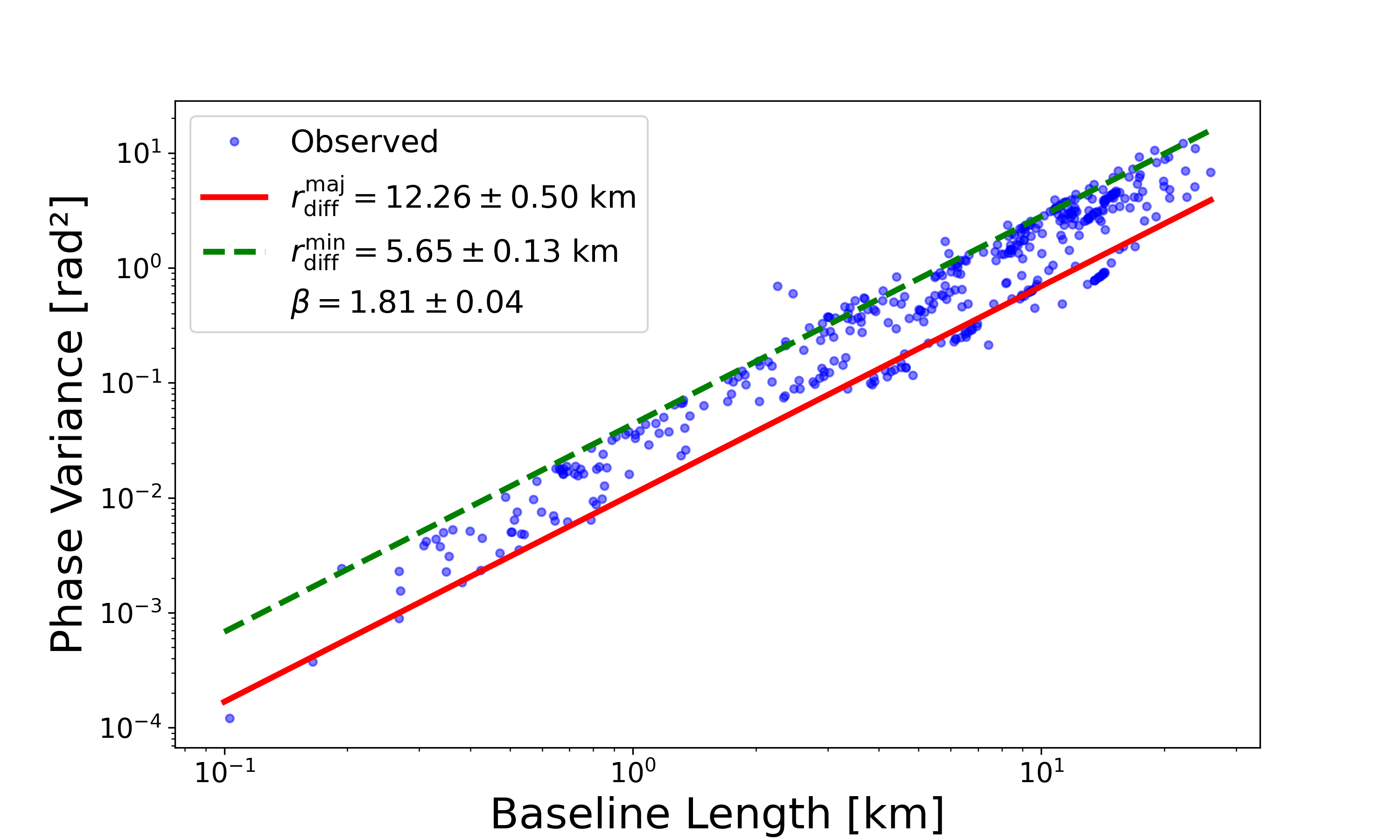}
    \hfill
    \includegraphics[width=0.49\linewidth]{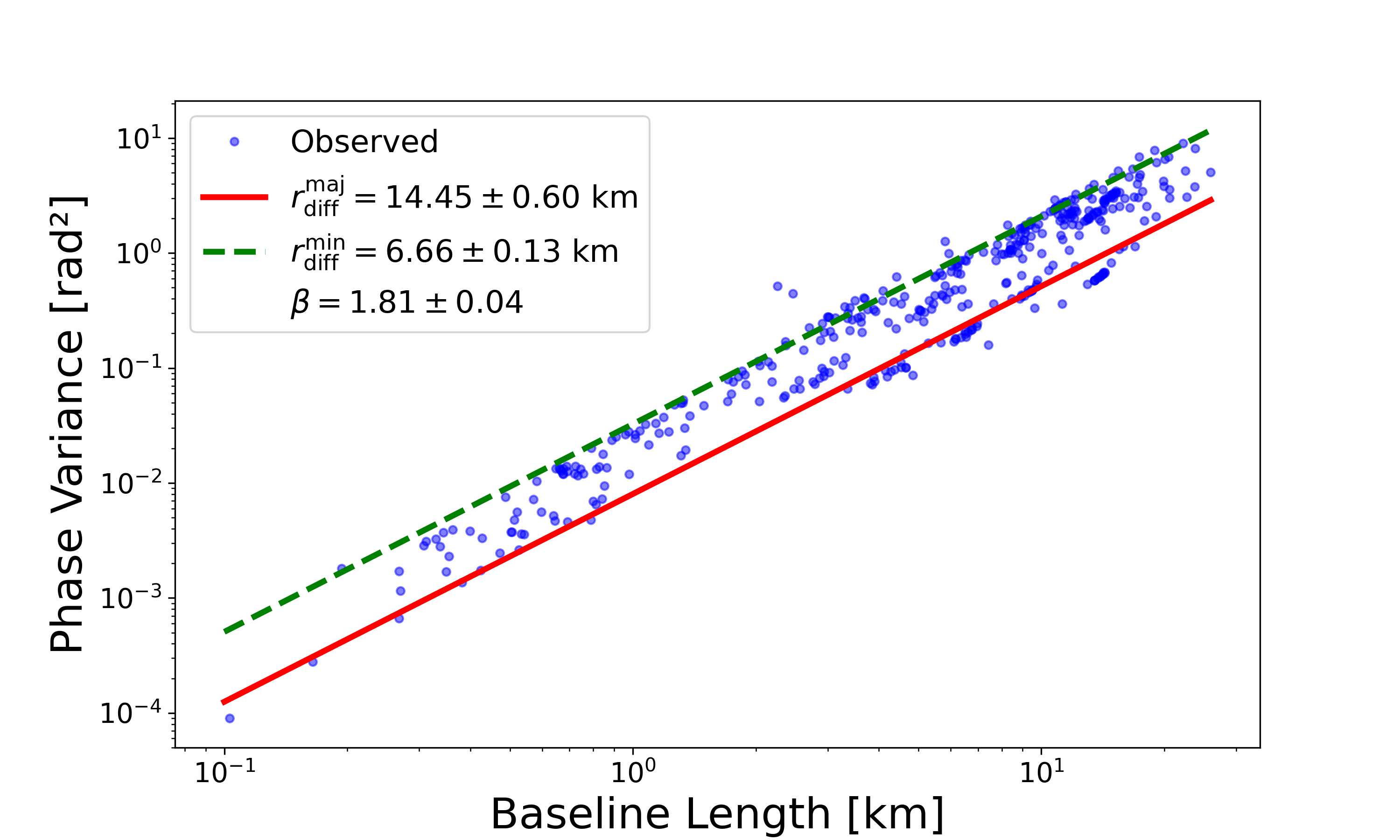}
    \caption{Two-dimensional phase structure function showing anisotropy. Left: At 625 MHz. Right: At 725 MHz. The green dotted and red solid lines indicate the fitted major and minor diffractive axes, respectively, based on the anisotropic structure function model (Eqn.~\ref{eq:2DKolmogorov}).}
    \label{fig:anisotropic_comparison}
\end{figure}

\begin{figure}[ht]
    \centering
    \includegraphics[width=0.49\linewidth]{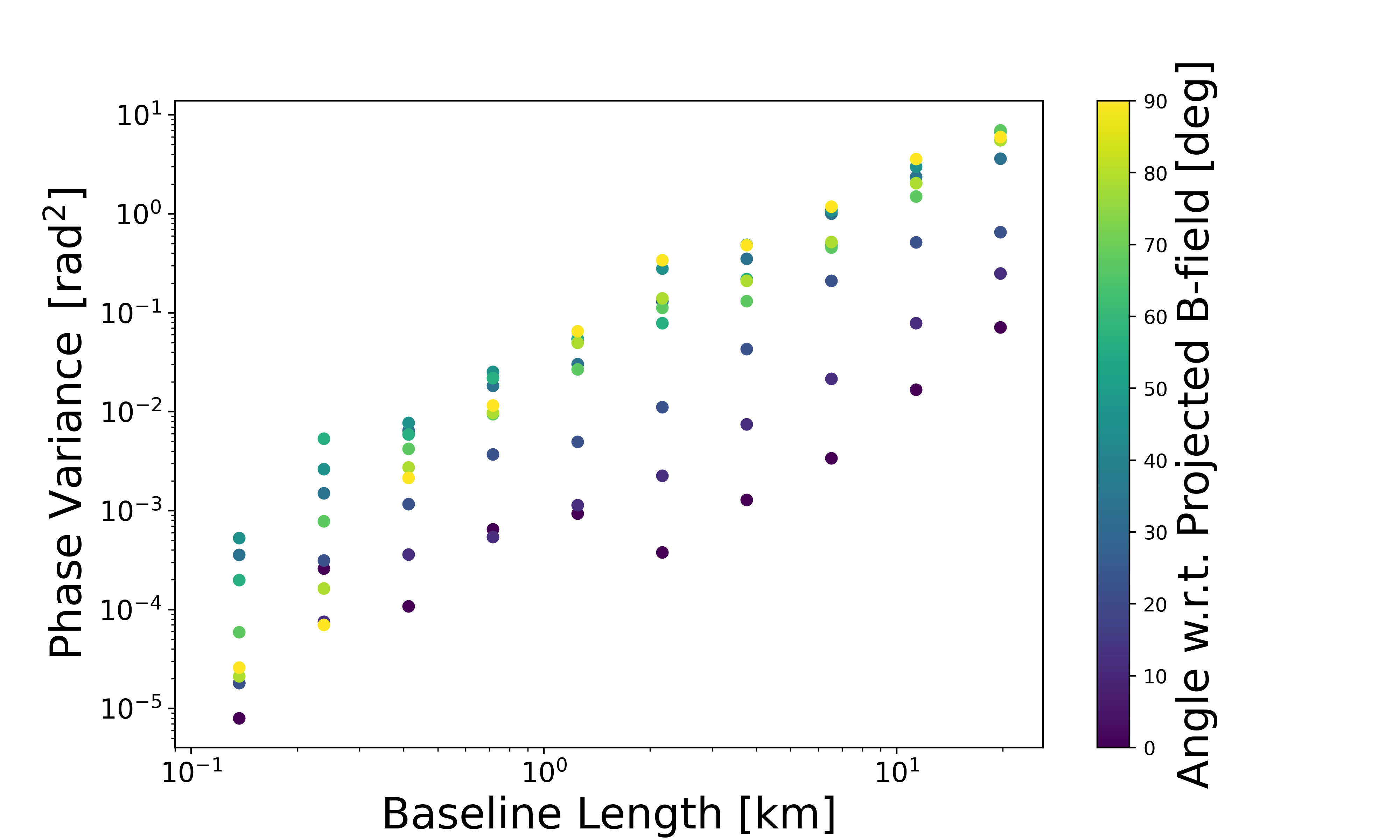}
    \hfill
    \includegraphics[width=0.49\linewidth]{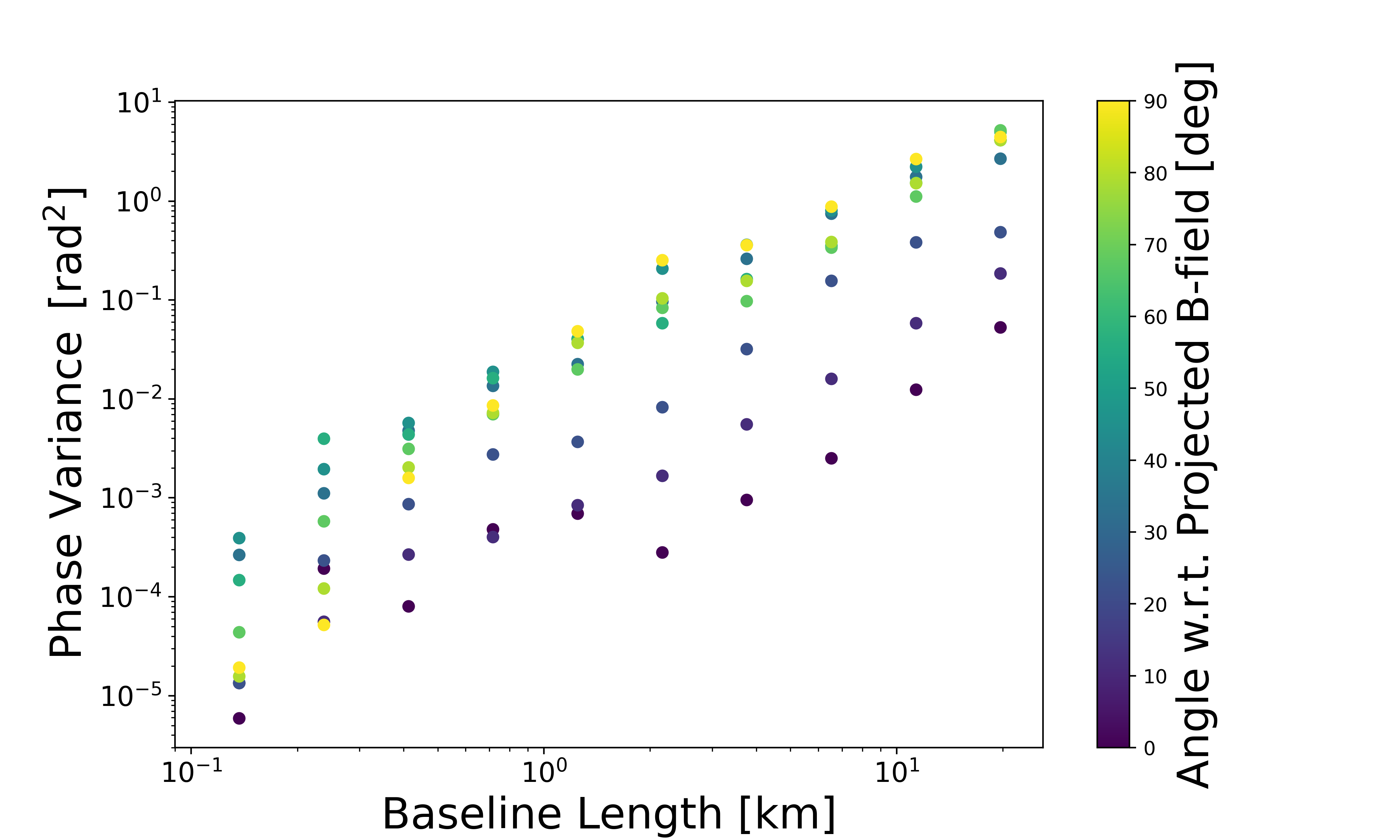}
    \caption{Phase structure function binned by angle relative to the projected Earth magnetic field. Left: At 625 MHz. Right: At 725 MHz. The colour bar indicates the angle in degrees.}
    \label{fig:phase_variance_Bangle_comparison}
\end{figure}

We selected three sub-bands with minimal RFI for analysis: 575--600, 600--625, and 700--725 MHz. We show the 600~MHz band as representative in the main text, with others in \ref{append1} to avoid repetition while demonstrating frequency trends.
Figures~\ref{fig:structure_func_comp},  \ref{fig:anisotropic_comparison}, and \ref{fig:phase_variance_Bangle_comparison} illustrate the results for two representative frequencies: 625 MHz and 725 MHz.

Here, we present the 1D phase structure functions with power-law fits (Fig.~\ref{fig:structure_func_comp}), the 2D anisotropic structure functions revealing turbulence directionality (Fig.~\ref{fig:anisotropic_comparison}), and the binned phase variance as a function of baseline orientation relative to the Earth's projected magnetic field (Fig.~\ref{fig:phase_variance_Bangle_comparison}).

Higher frequencies show improved data quality (less RFI flagging and better S/N), yielding more stable phase solutions. The diffractive scale increases with frequency ($r_{\mathrm{diff}} \sim 6.7\,\mathrm{km}$ at $600\,\mathrm{MHz}$ to $r_{\mathrm{diff}} \sim 8.33\,\mathrm{km}$ at $725\,\mathrm{MHz}$), as expected since phase fluctuations scale $\propto 1/\nu$, reducing sensitivity to small-scale TEC structure at higher bands.

\bibliographystyle{elsarticle-harv} 
\bibliography{ref}

\end{document}